\pdfoutput=1

\documentclass[11pt,twoside,a4paper,cmspaper,final,collab]{cms-tdr}
\def\svnVersion{375178}\def\cmsCernNoTag{CERN-EP/2016-118}\def\cmsCernDate{\today}\def\cmsMessage{Published in Physics Letters B as \href{http://dx.doi.org/10.1016/j.physletb.2016.10.007}{\doi{10.1016/j.physletb.2016.10.007.}}}

\begin{document}\cmsNoteHeader{TOP-13-008}

\hyphenation{had-ron-i-za-tion}
\hyphenation{cal-or-i-me-ter}
\hyphenation{de-vices}
\RCS$Revision: 370466 $
\RCS$HeadURL: svn+ssh://svn.cern.ch/reps/tdr2/papers/TOP-13-008/trunk/TOP-13-008.tex $
\RCS$Id: TOP-13-008.tex 370466 2016-10-12 09:30:35Z mara $
\newlength\cmsFigWidth
\ifthenelse{\boolean{cms@external}}{\setlength\cmsFigWidth{0.85\columnwidth}}{\setlength\cmsFigWidth{0.4\textwidth}}
\ifthenelse{\boolean{cms@external}}{\providecommand{\cmsLeft}{top}}{\providecommand{\cmsLeft}{left}}
\ifthenelse{\boolean{cms@external}}{\providecommand{\cmsRight}{bottom}}{\providecommand{\cmsRight}{right}}
\cmsNoteHeader{TOP-13-008}

\title{Measurement of the W boson helicity fractions in the decays of top quark pairs to lepton+jets final states produced in pp collisions at \texorpdfstring{$\sqrt{s}=8\TeV$}{sqrt(s) = 8 TeV}}

\date{\today}

\abstract{
The $\PW$ boson helicity fractions from top quark decays in  $\ttbar$  events are measured using data from proton-proton collisions at a centre-of-mass energy of 8\TeV. The data were collected in 2012 with the CMS detector at the LHC,  corresponding to an integrated luminosity of 19.8\fbinv. Events are reconstructed with either one muon or one electron, along with four jets in the final state, with two of the jets being identified as originating from b quarks. The measured helicity fractions from both channels are combined, yielding $F_\mathrm{0}=0.681\pm0.012\stat\pm 0.023\syst$,  $F_\mathrm{L}=0.323\pm 0.008\stat\pm 0.014\syst$, and  $F_\mathrm{R}=-0.004\pm 0.005\stat\pm 0.014\syst$ for the longitudinal, left-, and right-handed components of the helicity, respectively. These measurements of the $\PW$ boson helicity fractions are the most accurate to date and they agree with the predictions from the standard model.
}

\hypersetup{%
pdfauthor={CMS Collaboration},%
pdftitle={Measurement of the W boson helicity fractions in the decays of top quark pairs to lepton+jets final states produced in pp collisions at sqrt(s)=8 TeV},%
pdfsubject={CMS},%
pdfkeywords={CMS, physics, top, anomalous couplings,  helicity}}

\newcommand{\costh}{\ensuremath{\cos \theta^*}\xspace}
\newcommand{\costhhad}{\ensuremath{\cos^\text{had} \theta^*}\xspace}

\maketitle

\section{Introduction}
\label{sec:intro}

The data from proton-proton (pp) collisions produced at the CERN LHC
provide an excellent environment to investigate properties of the top quark, in the context of its production and  decay, with unprecedented precision.
Such measurements enable rigorous tests of the standard model (SM), and deviations from the SM predictions would indicate signs of possible new physics \cite{eqSaav3,np1,np2,np3}.

In particular, the $\PW$ boson helicity fractions in top quark decays are very sensitive to the Wtb vertex structure.
The $\PW$ boson helicity fractions are defined as the partial decay rate for a given helicity state divided by the total decay rate: $F_\mathrm{L,R,0}\equiv \Gamma_\mathrm{L,R,0}/\Gamma$, where $F_\mathrm{L}$, $F_\mathrm{R}$, and $F_\mathrm{0}$ are the left-handed, right-handed, and longitudinal helicity fractions, respectively.
 The helicity fractions are expected to be   $F_\mathrm{0}=0.687 \pm 0.005$, $F_\mathrm{L}=0.311 \pm 0.005$, and $F_\mathrm{R}=0.0017 \pm  0.0001$ at next-to-next-to-leading order (NNLO) in the SM,
including electroweak effects,  for a top quark mass $m_{\PQt} =172.8 \pm 1.3\GeV$~\cite{Czarnecki:2010gb}.
Anomalous Wtb couplings, i.e. those that do not arise in the SM, would alter these values.

\begin{figure*}[h]
\centerline{ \includegraphics[width = 0.5\textwidth]{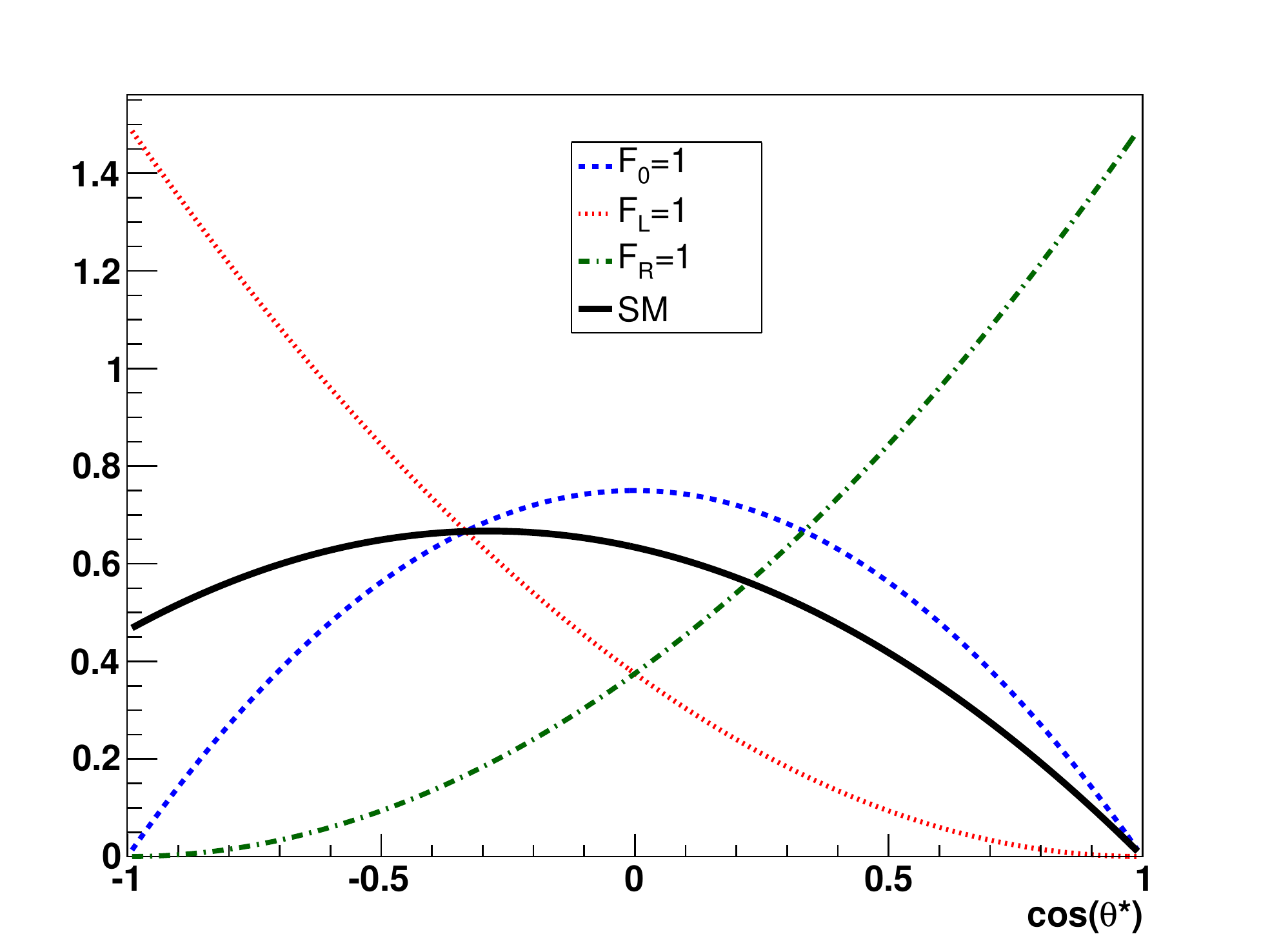} }
\caption{Predicted cos$\theta^*$ distributions for the different helicity fractions.
The distributions for the fractions $F_\mathrm{0},\ F_\mathrm{L}$, and $F_\mathrm{R}$
are shown as dashed, dotted, and dash-dotted lines, respectively, and the sum of the three contributions according to the SM predictions is displayed as a solid line. \label{fig:theory} }
\end{figure*}

Experimentally, the $\PW$ boson helicity can be measured through the study of angular distributions of the top quark decay products.
The helicity angle $\theta^*$ is defined as the angle between the direction of  either the down-type quark or the charged lepton  arising from the $\PW$ boson decay and
the reversed direction of the top quark, both in the rest frame of the $\PW$ boson.
The distribution for the cosine of the helicity angle depends on the helicity fractions in the following way,
\begin{linenomath}
\begin{equation}
\ifthenelse{\boolean{cms@external}}
{
\begin{split}
\frac{1}{\Gamma}\frac{\rd\Gamma}{\rd\cos{\theta^*}} = &  \phantom{+}\ \, \frac{3}{8}\left(1-\cos{\theta^*}\right)^2 F_\mathrm{L} \\
 & +  \frac{3}{4} (\sin{\theta^*} )^2 F_\mathrm{0} \\
 & +  \frac{3}{8}\left(1+\cos{\theta^*}\right)^2 F_\mathrm{R}.
\end{split}
}
{
\frac{1}{\Gamma}\frac{\rd\Gamma}{\rd\cos{\theta^*}} =
\frac{3}{8}\left(1-\cos{\theta^*}\right)^2 F_\mathrm{L} +
\frac{3}{4} (\sin{\theta^*} )^2 F_\mathrm{0} +
\frac{3}{8}\left(1+\cos{\theta^*}\right)^2 F_\mathrm{R}.
}
\label{eq:costhetastar}
\end{equation}
\end{linenomath}
This dependence is shown in Fig. \ref{fig:theory} for each contribution separately, normalised to unity, and for the SM expectation. Charged leptons (or down-type quarks) from left-handed W bosons are preferentially emitted in the opposite direction of the W boson,
and thus tend to have lower momentum and be closer to the b jet from the top quark decay,
as compared to charged leptons (or down-type quarks) from longitudinal or right-handed W bosons.

The measurement of the $\PW$ boson helicity is sensitive to the presence of non-SM couplings between the $\PW$ boson, the top quark, and the bottom quark.
A general parametrisation of the Wtb vertex can be expressed as~\cite{eqSaav1,eqSaav3}
\begin{linenomath}
\begin{equation}
\ifthenelse{\boolean{cms@external}}
{
\begin{split}
\mathcal{L}_\mathrm{Wtb}  = &  - \frac{g}{\sqrt 2} \bar{b}  \gamma^{\mu} ( V_\mathrm{L} P_\mathrm{L} + V_\mathrm{R} P_\mathrm{R} ) t W_\mu^-  \\
 & - \frac{g}{\sqrt 2} \bar {b}  \frac{i \sigma^{\mu \nu} q_\nu}{M_\PW}  \left( g_\mathrm{L} P_\mathrm{L} + g_\mathrm{R} P_\mathrm{R} \right) t\; W_\mu^- + \mathrm{h.c.} ,
\end{split}
}
{
\mathcal{L}_\mathrm{Wtb}  =  - \frac{g}{\sqrt 2} \bar{b}  \gamma^{\mu} ( V_\mathrm{L} P_\mathrm{L} + V_\mathrm{R} P_\mathrm{R} ) t W_\mu^-
 - \frac{g}{\sqrt 2} \bar {b}  \frac{i \sigma^{\mu \nu} q_\nu}{M_\PW}  \left( g_\mathrm{L} P_\mathrm{L} + g_\mathrm{R} P_\mathrm{R} \right) t\; W_\mu^- + \mathrm{h.c.} ,
 }
\label{eq:Wtb0}
\end{equation}
\end{linenomath}
where $V_\mathrm{L},\ V_\mathrm{R},\ g_\mathrm{L},\ g_\mathrm{R}$ are vector and tensor couplings (complex constants), $q=p_{\PQt}-p_\PQb$, and $p_{\PQt}$ ($p_\PQb$) is the four-momentum of the top quark (b quark),  $P_\mathrm{L}\ (P_\mathrm{R})$ is the left (right) projection operator, and h.c.\ denotes the Hermitian conjugate.
Hermiticity conditions on the possible dimension-six Lagrangian terms also impose $\mathrm{Im}(V_\mathrm{L})=0$~\cite{eqSaav2}.
In the SM and at tree level, \mbox{$V_\mathrm{L} = V_\mathrm{tb}$},  where $V_\mathrm{tb} \approx 1$ is the Cabibbo--Kobayashi--Maskawa matrix element connecting the top and the bottom quarks and
$V_\mathrm{R}=g_\mathrm{L}=g_\mathrm{R}=0$.
The relationships between the $\PW$ boson helicity fractions and the anomalous couplings including dependences on the b quark mass
 are given in Ref.~\cite{SaavedraBernabeu}.

The helicity fractions of $\PW$ bosons in top quark decays were first measured at the Tevatron Collider \cite{d0new,cdfFull,TevComb}. They have been also measured at the LHC, using samples containing \ttbar~events obtained in pp collisions at 7\TeV, and having either one \cite{AtlasPaper,TOP-11-020} or two \cite{AtlasPaper} charged leptons in the final state.
The CMS Collaboration also reported measurements using event topologies that contain one single reconstructed top quark \cite{singtop}, in pp collisions at 8\TeV.
Limits on anomalous couplings have also been reported, derived from W boson helicity measurements  \cite{AtlasPaper,TOP-11-020,singtop}, and  from  single top quark differential cross section
production measurements \cite{AnomATLAS}.

This Letter describes a measurement of the $\PW$ boson helicity fractions in  \ttbar events involving one lepton and multiple jets, $\ttbar\to (\PW^+\PQb) \ (\PW^-\bar\PQb) \ \to \ (\ell^+\nu_\ell\PQb) \  (\PQq\PAQq^\prime\PAQb)$,
and its charge conjugate, where $\ell$ is an electron or a muon,  including those from leptonic decays of a tau lepton.
Final states corresponding to such processes are referred to
as lepton+jets.  The measurement relies on the analysis strategy described in  Ref. \cite{TOP-11-020}.
The measurement is performed using pp collisions at centre-of-mass energy of 8\TeV, corresponding to an integrated luminosity of  $19.8\fbinv$, collected during 2012 by the CMS detector.

\section{The CMS detector}

The CMS detector is a multipurpose apparatus of cylindrical design with respect to the proton beams.
The main features of the detector relevant for this analysis are briefly described here.
Charged particle trajectories are measured by a silicon pixel and strip tracker, covering the pseudorapidity range  $|\eta| < 2.5$.
The inner tracker is immersed in a 3.8\unit{T} magnetic field provided by a superconducting solenoid of 6\unit{m} in diameter that also encompasses several calorimeters.
A lead tungstate crystal electromagnetic calorimeter (ECAL), and a brass and scintillator hadronic calorimeter  surround the tracking volume and cover the region $|\eta| < 3$.
 Quartz fibre and steel hadron forward calorimeters extend the coverage to $|\eta| \le 5$.
Muons are identified in gas ionisation detectors embedded in the steel return yoke of the magnet.
The data for this analysis are recorded using a two-level trigger system.
A more detailed description of the CMS detector, together with a definition of the coordinate system used and the relevant kinematic variables, can be found in Ref.~\cite{CMS_detector}.

\section{Data and simulated samples}

Signal events
corresponding to top quark pairs that decay to lepton+jets final states are expected to
contain one isolated lepton (electron or muon) together with at least four
jets, two of which originate from b quark fragmentation. Such events are
referred to separately
as $\Pe$+jets or $\mu$+jets, respectively, or when combined as $\ell$+jets.
Background events containing a single isolated lepton and four reconstructed jets arise mainly  from processes that produce events containing a
single top quark, processes that produce multijet events in association with a $\PW$ boson that decays leptonically ($\PW$+jets), or Drell--Yan processes accompanied by multiple jets (DY+jets) when one of the leptons is misidentified as a jet or goes undetected.
Multijet processes can also mimic lepton+jets final states, if a
 jet is reconstructed as an electromagnetic shower or, more unlikely, if a nonprompt muon from a hadron decay in flight fulfils all identification criteria of a prompt muon.

 Simulated  Monte Carlo (MC) samples,  interfaced with \GEANTfour \cite{geant},
are used to account for detector resolution and acceptance effects, as well as to estimate the contribution from
background processes that have  characteristics similar to lepton+jets
final states in \ttbar~decays.
A signal \ttbar sample, which also provides a reference for the SM (see Eq. (\ref{eq:Reweighiting:weightDef})), is simulated using \MADGRAPH v5.1.3.30 \cite{MADGRAPH} with matrix elements having up to three extra partons in the final state.
The parton distribution function (PDF) set {\sc CTEQ6L1}~\cite{cteq} is used when simulating
this reference \ttbar~sample.
The \MADGRAPH generator is interfaced with \PYTHIA v6.426 \cite{PYTHIA}, tune Z2* \cite{tuneZ2},  to simulate hadronisation and parton fragmentation, and also with \TAUOLA v27.121.5 \cite{tauola} to simulate $\tau$ lepton decays.
This SM reference \ttbar sample is simulated assuming
$m_{\PQt}=172.5\GeV$, which results in the following leading-order (LO) $\PW$ boson helicity fractions for that sample:
\begin{linenomath}
 \begin{equation}
F^\mathrm{SM}_\mathrm{0}=0.6902,\ \ F^\mathrm{SM}_\mathrm{L}=0.3089,\ \ F^\mathrm{SM}_\mathrm{R}=0.0009.
 \label{eq:refhel}
 \end{equation}
\end{linenomath}
Single top quark events in the $s$, $t$, and tW
channels are generated using \POWHEG  v1.0~\cite{POWHEG} and \PYTHIA  interfaced with  \TAUOLA, with the PDF set {\sc CTEQ6M}~\cite{cteq}.
Background $\PW$+jets and DY+jets processes are simulated using \MADGRAPH with the PDF set {\sc CTEQ6L1}, followed by \PYTHIA for fragmentation and hadronisation.
Finally, background multijet processes  are simulated using the \PYTHIA event generator.

Corrections are applied to the simulated samples so that resolutions, energy scales, and efficiencies as functions of  \pt and $\eta$ of jets
\cite{ref:jets} and leptons \cite{diff2}  measured in data are well described.
The effect of multiple pp collisions occurring in the same bunch crossing (pileup) is also taken into account in the simulation.

The data samples selected for this measurement were recorded  using inclusive single-lepton triggers, which require at least one isolated electron (muon) with $\pt>27\ (24)$\GeV, used to define the $\Pe$+jets ($\mu$+jets) data sample.

The decay products of candidate top quarks are reconstructed
 using the CMS particle-flow (PF) algorithm, described in detail elsewhere~\cite{CMS-PAS-PFT-09-001,CMS-PAS-PFT-10-001}.
Individual charged particles identified as coming from pileup interactions are removed from the event.
Effects of neutral particles from pileup interactions are mitigated by applying corrections based on event properties.
Leptons are required to originate from the primary vertex of the event~\cite{ref:TRK-11-001}.
A lepton is determined to be isolated using
a variable  computed as the
total transverse momentum of all particles (except the lepton itself) contained within a cone of radius 0.4,
centred on the lepton direction, relative to the transverse momentum of the lepton.
Electrons are identified by using a multivariate analysis (MVA)~\cite{ref:electrons} based on information from the inner tracker and the ECAL.
Events are selected for the $\Pe$+jets data sample if the identified electron has an MVA discriminant value greater than 0.9,
 is determined to be isolated,  has  $\pt>30\GeV$, and $|\eta|<2.5$.
Muons are identified
 by matching information from the inner tracker and the muon spectrometer~\cite{ref:muons}.
Events are selected for the $\mu$+jets data sample if they contain an isolated muon,
 $\pt>26$\GeV, and $\abs{\eta} < 2.1$.
Events with at least one additional isolated electron or muon are vetoed to reject backgrounds from dileptonic decay modes of \ttbar and DY+jets processes.
Jets are reconstructed \cite{ref:jets} using the anti-$\kt$ clustering algorithm \cite{antikt_jet_algo}, with a distance parameter of 0.5.
The selected or vetoed leptons described above are not allowed to be clustered into jets, to avoid ambiguities.

The event selection requires at least four reconstructed jets having $|\eta|<2.4$,
of which the four most energetic jets are required to have $\pt$ higher than 55, 45, 35, and 20\GeV.
 Events with additional jets are not vetoed.  The transverse momentum imbalance of the event $\vec{p}_\mathrm{T}^\mathrm{miss}$ is determined by summing the
negative transverse momentum over
 all reconstructed particles, excluding those charged
particles not associated with the primary vertex.

The transverse mass of the $\PW$ boson is defined as $M_\mathrm{T} = \sqrt{\smash[b]{ 2 p^\ell_\mathrm{T} p_\mathrm{T}^\mathrm{miss}(1-\cos(\Delta\phi))}}$, where $p^\ell_\mathrm{T}$ is the transverse momentum of the lepton, $p_\mathrm{T}^\mathrm{miss}$ is the magnitude of $\vec{p}_\mathrm{T}^\mathrm{miss}$, and $\Delta\phi$ is the angle in the $(x,y)$ plane between the direction of the lepton and $\vec{p}_\mathrm{ T}^\mathrm{miss}$.
To reduce the  multijet background and suppress dilepton events from \ttbar~processes, events are required to have $30 < M_\mathrm{T} < 200\GeV$.
All backgrounds are further suppressed by requiring that at least two jets be identified as originating from b quarks. All jets
with $p_\mathrm{T}>20\GeV$ are considered as b quark candidates, including those that are not among the four most energetic.

The combined secondary vertex algorithm  \cite{ref:btag,btagpas} tags
b quark jets with an efficiency of about 70\% and mistags jets originating from gluons, u, d, or s quarks with a probability of about 1\%,  for the typical $p_\mathrm{T}$ ranges (30--100 GeV) probed in \texorpdfstring{\ttbar}{t-tbar} events.
 Charm jets have a probability of $\approx$20\% of being tagged as b quark jets.
The residual multijet backgrounds, already strongly suppressed by the b tagging requirement described above, are estimated by normalising simulated event samples to yields in control data samples.
The control samples are defined by selection criteria which are similar to those for the signal, but which have no b tagging requirement, and have $ M_\mathrm{T} < 30\GeV$ for the $\mu$+jets channel or have an electron
MVA discriminant value smaller than 0.5 for the $\Pe$+jets channel.
The estimated amount of multijet events is $\approx$2\% of the $\Pe$+jets sample, and less than 1\% of the $\mu$+jets sample.
The contributions of all other residual backgrounds are determined using simulation.

\section{Reconstruction of the \texorpdfstring{\ttbar}{t-tbar} system and reweighting method}

The reconstruction of the \ttbar system, described in detail in Ref.  \cite{TOP-11-020}, relies on testing the selected lepton,
the measured $\ptvecmiss$, and all selected jets for their compatibility with the top quark decay products from the leptonic
 $(\PQt \to \PQb \PW\ \to \PQb \ell \nu)$ and hadronic $(\PQt \to \PQb \PW\to \PQb \PQq \PAQq^\prime)$ branches.
The unmeasured component of the neutrino momentum $p^\nu_z$ is determined by
 requiring the \ttbar system to be consistent with the invariant masses of two top quarks and two $\PW$ bosons.
 With these constraints,
\PQb jets are correctly assigned to the leptonic (hadronic) branch in about 74\% (71\%) of signal events.
After the assignment, a kinematic fit is  performed, where the momenta of the measured
  jets and lepton are allowed to vary within their resolutions to better comply with the mass constraints, leading to an improved determination of $p^\nu_z$ and a more accurate reconstruction of the $\ttbar$ system.
In about 5--7\% of the selected events, the fit fails to find a solution that is compatible with the constraints and such events are discarded.
The number of data events passing all selection criteria, including the fit convergence, is $71\,458$ in the $\Pe$+jets sample and 70 986 in the  $\mu$+jets sample.
A study using simulations normalised to the most precise theoretical cross sections available to date \cite{ttbarNLO1,ttbarNLO2,singlet8Tev,fewz} indicates that the final sample composition is largely dominated by \ttbar~events, with about 82\%  of events
from the $\ell$+jets decay mode, ${\approx}10\%$ from other decay modes (including $\tau$ leptons), and ${\approx}3.5\%$ of the events
 from single top quark processes.
The remaining events come from backgrounds not containing top quarks in the final state.

The  method~\cite{TOP-11-020} employed to measure the $\PW$ boson helicity fractions $(F_\mathrm{L},F_\mathrm{0},F_\mathrm{R})\equiv \vec{F}$ consists of maximising a binned Poisson likelihood function constructed using the number of observed events in data ${N_\text{data}}(i)$ and expected events from MC simulation ${ N_\mathrm{MC}}(i;\vec{F})$,
 in each bin $i$ of the reconstructed $\cos\theta^*_\mathrm{rec}$ distribution,
\begin{linenomath}
 \begin{equation}
\mathcal{L}(\vec{F}) = \prod_{i} \frac{N_\mathrm{MC}(i;\vec{F})^{\displaystyle~N_\text{data}(i)}}{\Bigl[ N_\text{data}(i)\Bigr]!}~\exp{[-N_\mathrm{MC}(i;\vec{F})]}.
 \end{equation}
\end{linenomath}
\noindent
While the charged lepton  is easily identified  in the leptonic branch of \ttbar~decays, the down-type quark jet  arising from the $\PW$ boson decay in the hadronic branch of \ttbar~decays can not be experimentally distinguished from the up-type quark jet.
Due to this ambiguity, only the absolute value $|\cos\theta^*_\text{rec}|$ can be reconstructed for the hadronic branch.
Hence, only the leptonic branch measurement of $\cos\theta^*_\text{rec}$  is used in this analysis.
The expected numbers of events from background processes, ${N_{\PW+\text{jets}}}(i)$, ${N_{\mathrm{DY}+\text{jets}}}(i)$, and ${N_\text{multijet}}(i)$ represent $\PW$ boson production in association with multiple jets, Drell--Yan production in association with multiple jets, and production of multiple jets, which do not depend on the $\PW$ boson helicity fractions.
For the processes containing top quarks, the number of expected events in a given bin $i$ is modified by reweighting each event in that bin by
a factor $w$, defined for each decaying branch as
\begin{equation}
\ifthenelse{\boolean{cms@external}}
{
\begin{split}
w_\text{lep/had/single-\PQt}(\cos\theta^{*}_\text{gen}; & \vec{F})  \equiv \\
& \left[\  \phantom{+} \frac{3}{8} F_\mathrm{L} (1-\cos\theta^{*}_\text{gen})^2 \right. \\
& \left. \ + \ \frac{3}{4} F_\mathrm{0} \sin^{2}\theta^{*}_\text{gen}  \right. \\
& \left. \ + \  \frac{3}{8} F_\mathrm{R} (1+\cos\theta^{*}_\text{gen})^{2} \ \right] / \\
& \left[ \ \phantom{+} \frac{3}{8} F^\mathrm{SM}_\mathrm{L }(1-\cos\theta^{*}_\text{gen})^2 \right. \\
& \left. \ + \ \frac{3}{4} F^\mathrm{SM}_\mathrm{0 } \sin^{2}\theta^{*}_\text{gen} \right. \\
& \left. \ + \ \frac{3}{8} F^\mathrm{SM}_\mathrm{R } (1+\cos\theta^{*}_\text{gen})^{2} \ \right],
\end{split}
}
{
  w_\text{lep/had/single-\PQt}(\cos\theta^{*}_\text{gen};\vec{F}) \equiv
   \frac
  {\displaystyle
  \frac{3}{8} F_\mathrm{L} (1-\cos\theta^{*}_\text{gen})^2 +
  \frac{3}{4} F_\mathrm{0} \sin^{2}\theta^{*}_\text{gen} +
  \frac{3}{8} F_\mathrm{R} (1+\cos\theta^{*}_\text{gen})^{2}
  }
  {\displaystyle
  \frac{3}{8} F^\mathrm{SM}_\mathrm{L }(1-\cos\theta^{*}_\text{gen})^2 +
  \frac{3}{4} F^\mathrm{SM}_\mathrm{0 } \sin^{2}\theta^{*}_\text{gen} +
  \frac{3}{8} F^\mathrm{SM}_\mathrm{R } (1+\cos\theta^{*}_\text{gen})^{2}
  },
}
\label{eq:Reweighiting:weightDef}
\end{equation}
where $\theta^{*}_\text{gen}$ is the helicity angle (specified at matrix element level) of a particular decay branch,
and  $F^\mathrm{SM}_\mathrm{L}, F^\mathrm{SM}_\mathrm{0}, F^\mathrm{SM}_\mathrm{R}$ are given in Eq.~(\ref{eq:refhel}).
Therefore, the number of expected events, as a function of the helicity fractions to be measured, is
\begin{equation}
\ifthenelse{\boolean{cms@external}}
{
\begin{split}
{N_\mathrm{MC}}(i;\vec{F})  = & \phantom{+} \ \, {N_{\ttbar}}(i;\vec{F}) \\
& + {N_\text{single-\PQt}}(i;\vec{F}) \\
& + {N_{\PW+\text{jets}}}(i) \\
& + {N_{\mathrm{DY}+\text{jets}}}(i) \\
& + {N_\text{multijet}}(i),
\end{split}
}{
{N_\mathrm{MC}}(i;\vec{F})  =  {N_{\ttbar}}(i;\vec{F}) + N_\text{single-\PQt}(i;\vec{F}) + {N_\mathrm{\PW+jets}}(i)+{N_\mathrm{DY+jets}}(i)+{N_\mathrm{multijet}}(i),
}
 \end{equation}
where
\ifthenelse{\boolean{cms@external}}
{
\begin{equation}
\begin{split}
N_{\ttbar} (i;\vec{F}) =   \mathcal{ F}_{\ttbar} \left[ \sum_\text{\ttbar events in bin $i$}  \right. & \left. \ \ \ \ w_\mathrm{lep}(\cos\theta^{*}_\text{gen};\vec{F})  \right. \\
													  & \left. \times \ w_\text{had}(\cos\theta^{*}_\text{gen};\vec{F}) \ \ \right] , \\
N_\mathrm{single\textnormal{-}t}(i;\vec{F}) =  \sum_{\text{single-\PQt events in bin $i$}} & w_\text{single-\PQt}(\cos\theta^{*}_\text{gen};\vec{F})
\end{split}
\label{eq::Reweighting::WeightDef1}
\end{equation}
}
{
\begin{eqnarray}
{N_{\ttbar}}(i;\vec{F}) & = & \mathcal{ F}_{\ttbar}\left[\sum_\text{\ttbar events in bin $i$}  w_\mathrm{lep}(\cos\theta^{*}_\text{gen};\vec{F}) \times w_\mathrm{had}(\cos\theta^{*}_\text{gen};\vec{F}) \right]
\label{eq::Reweighting::WeightDef1}, \\
N_\mathrm{single\textnormal{-}t}(i;\vec{F}) & = & \sum_\text{single-\PQt events in bin $i$} w_\text{single-\PQt}(\cos\theta^{*}_\text{gen};\vec{F})
\end{eqnarray}
}

represent the expected number of events fulfilling event selection criteria  for processes involving top quark pair,   and single top quark production, respectively.
The normalisation factor $\mathcal{F}_{\ttbar}$ for the \ttbar sample is a single free parameter in the fit across all bins.
The expected cross section for the simulated reference \ttbar sample is $252.9^{+13.3}_{-14.5}$\unit{pb},  calculated at NNLO and next-to-next-to-leading-log (NNLL) accuracy \cite{ttbarNLO1,ttbarNLO2}, and describes the data well.
The fitted values of $\mathcal{F}_{\ttbar}$ in both $\Pe$+jets and $\mu$+jets channels
are compatible with 1.00 within 3\%.
The overall normalisation factor for simulated single top quark events is not modified in the fit and the uncertainty in the assumed cross section is considered as a source of systematic uncertainty.
Finally, the unitarity constraint ($F_\mathrm{L}+F_\mathrm{0}+F_\mathrm{R}=1$) is imposed, so that one of the helicity fractions, namely $F_\mathrm{R}$, is
bound by the measurement of the other two.

The method was validated using pseudo-experiments, where the fitting procedure was performed on pseudo-data, mimicking altered helicity fractions.
Linearity tests show that the fitting procedure correctly retrieves
the helicity fractions of altered input values for $F_0\in[0.50,0.85]$
 and $F_L\in[0.20,0.50]$. Likewise the corresponding
statistical uncertainties were verified using sets of statistically uncorrelated pseudo-data.

\begin{table*}[h]
\begin{center}
\topcaption{ Systematic uncertainties on the measurements of the $\PW$ boson helicity fractions
 from lepton+jets events. The cases in which the statistical precision of the limited sample size was assigned as systematic uncertainties are denoted by
the symbol (*).
\label{tab:system}}
\tabcolsep=0.15cm
\ifthenelse{\boolean{cms@external}}{}{\resizebox{\textwidth}{!}}
{
\begin{tabular}{lcc|c c |c c } \cline{2-7}
  & \multicolumn{2}{c|}{$\Pe$+jets}  & \multicolumn{2}{c|}{$\mu$+jets}  & \multicolumn{2}{c}{$\ell$+jets} \\ \cline{2-7}
 & $\pm$  $\Delta F_\mathrm{0}$  &  $\pm$ $\Delta F_\mathrm{L}$   & $\pm$  $\Delta F_\mathrm{0}$  &  $\pm$ $\Delta F_\mathrm{L}$   & $\pm$  $\Delta F_\mathrm{0}$  &  $\pm$ $\Delta F_\mathrm{L}$  \\ \hline
                     JES     & 0.004   & 0.003		  &   0.005  	& 0.003		      & 0.005		& 0.003  	   \\
                      JER    & 0.001   & 0.002		  &   0.004   	& 0.003  	      & 0.003		& 0.003  	   \\
             b tagging eff.  & 0.001   & ${<}10^{-3}$ 	  &   0.001   	& ${<}10^{-3}$        & 0.001		& ${<}10^{-3}$     \\
             Lepton eff.     & 0.001   & 0.002 		  &   0.001   	& 0.001  	      & 0.001		& 0.001  	  \\
         Single top normal.  & 0.002   & ${<}10^{-3}$     &   0.003    	& 0.001               & 0.003 	& 0.001  	   \\
              $\PW$+jets bkg.  & 0.008   & 0.001 	          &   0.007    	& 0.001               & 0.007 	& 0.001  	   \\
               DY+jets bkg.  & 0.002   & ${<}10^{-3}$ 	  &   0.001    	& ${<}10^{-3}$       &    0.001 	& ${<}10^{-3}$      \\
    Multijet bkg.            & 0.023   & 0.007            &   0.007          & 0.003              & 0.008 	& 0.001  	   \\
                      Pileup & 0.001   & 0.001 		  &   ${<}10^{-3}$     & ${<}10^{-3}$           & 0.001		& ${<}10^{-3}$     \\
            Top quark mass   & 0.012   & 0.008  	  &   0.010 (*)    	& 0.008 (*)               & 0.010 	& 0.007     \\
              \ttbar~scales  & 0.011   & 0.008  (*)	  &   0.014    	& 0.007  (*)              & 0.012 	& 0.007     \\
         \ttbar~match. scale & 0.011 (*)   & 0.007 (*)  	  &   0.010    	& 0.007                & 0.009 	& 0.007     \\
 \ttbar~MC and hadronisation & 0.015   & 0.009            &   0.005    	& 0.003               & 0.006 	& 0.004  	  \\
         \ttbar \pt reweight & 0.011   & 0.010  	  &   ${<}10^{-3}$   	& 0.001        	   	& ${<}10^{-3}$  	& 0.002  	  \\
             Limited MC size & 0.002   & 0.001 	          &   0.002      & 0.001        	& 0.002 	& 0.001  	  \\
                     PDF     & 0.004   & 0.001 		  &   0.002      & 0.001       	& 0.002 	& 0.001  	   \\ \hline
                Total        & 0.037   & 0.020 		  &  0.024   & 0.014  & 0.023 & 0.014 \\ \hline
\end{tabular}
}
\end{center}
\end{table*}

\begin{figure*}[h]
\centerline{ \includegraphics[width = 0.5\textwidth]{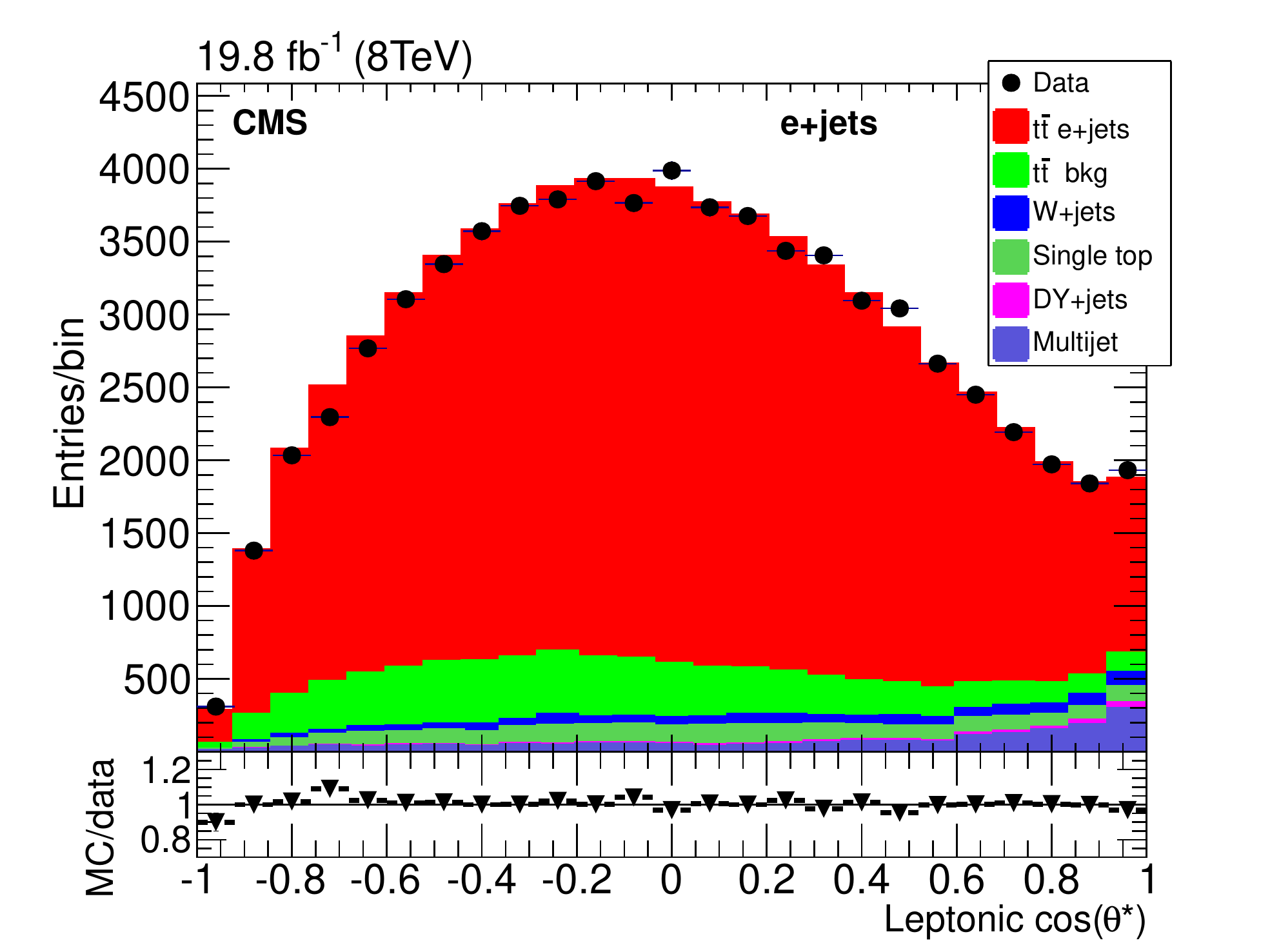}  \includegraphics[width = 0.5\textwidth]{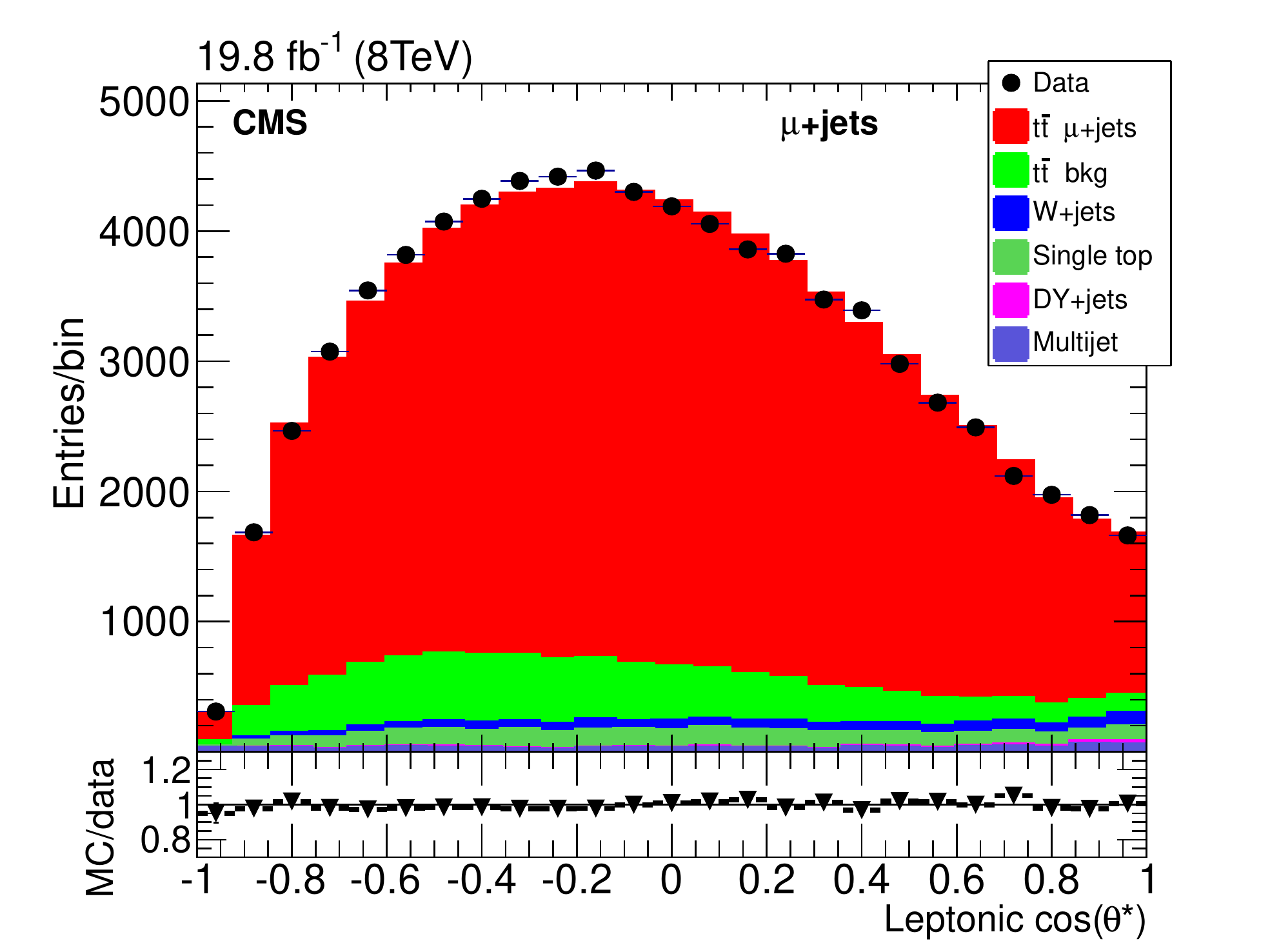}}
\centerline{ \includegraphics[width = 0.5\textwidth]{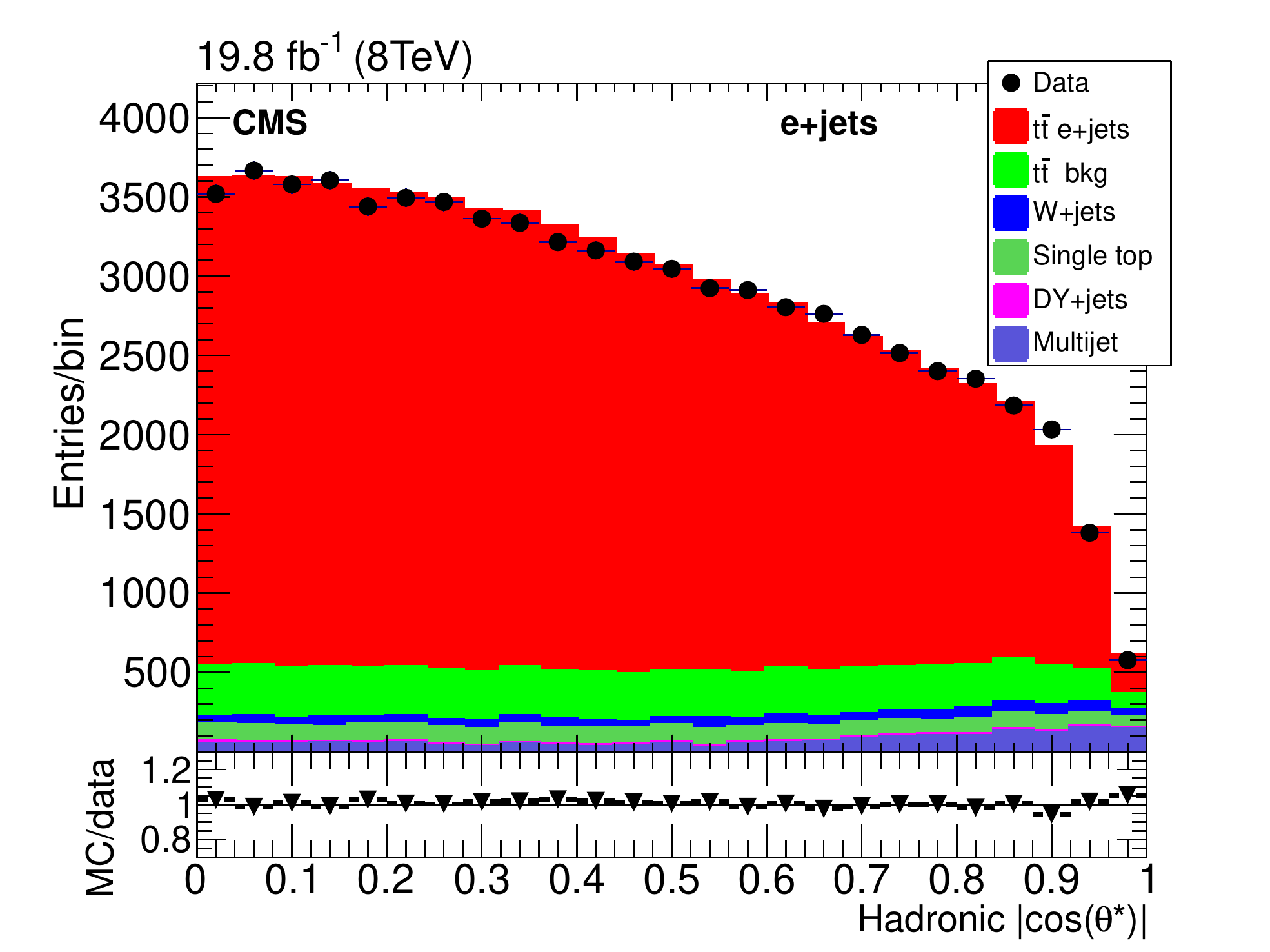}  \includegraphics[width = 0.5\textwidth]{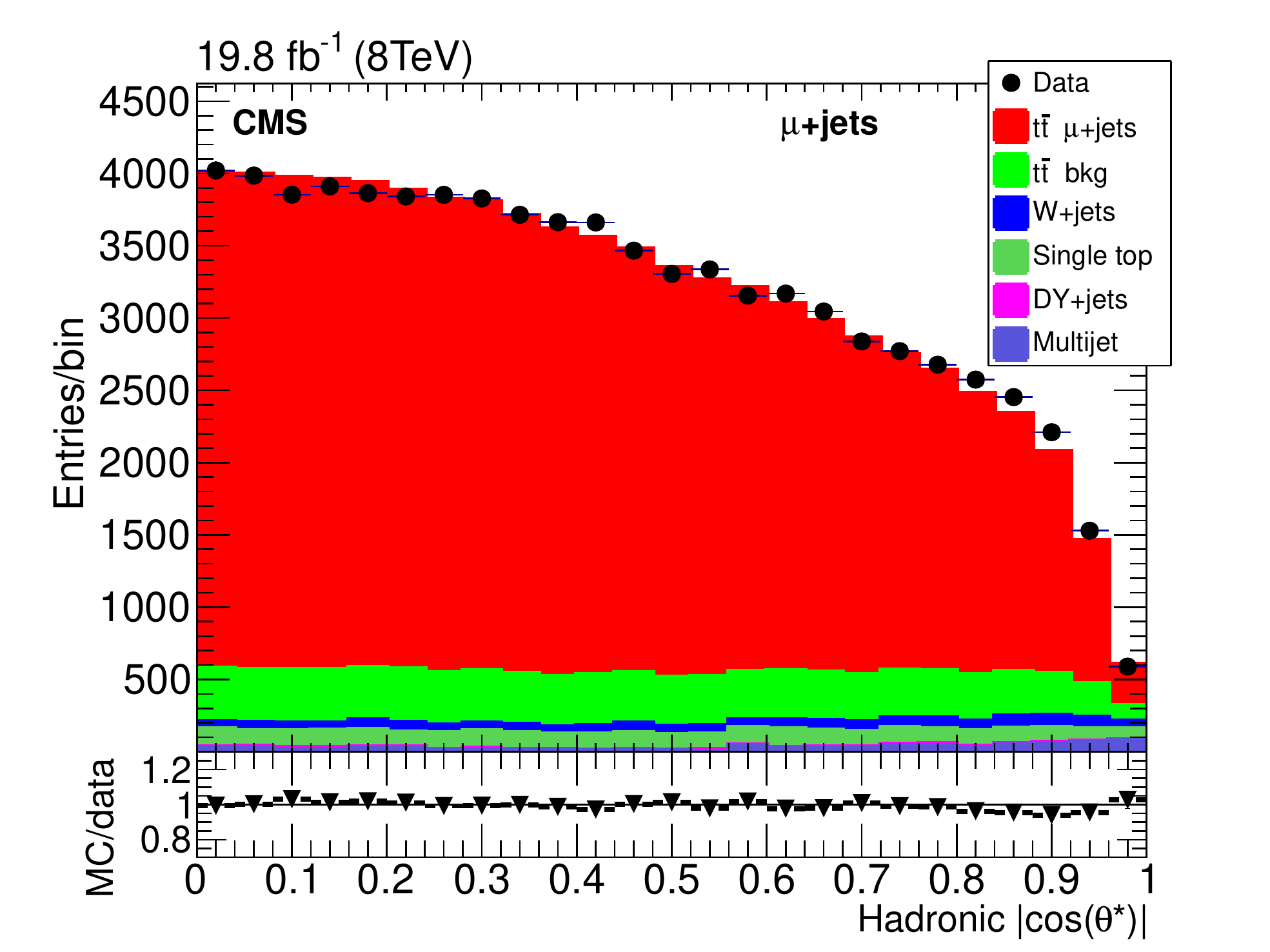}}
\caption{
Distributions for the cosine of the helicity angle in the leptonic (upper row)  and hadronic (lower row) branches, for the $\Pe$+jets (left) and $\mu$+jets (right) decay channels.
The combined $\ell$+jets post-fit measurements of the helicity fractions were used in the simulation of \ttbar~and single top quark events.
The data are displayed as solid points, simulated samples of \ttbar (signal) processes
and the contribution from background processes as histograms.
At the bottom of each plot, the ratio between MC simulation and data is displayed. The error bars correspond to the statistical uncertainties. \label{fig2}
}
\end{figure*}

\begin{figure*}[h]
\begin{center}
\centerline{ \includegraphics[scale=0.45]{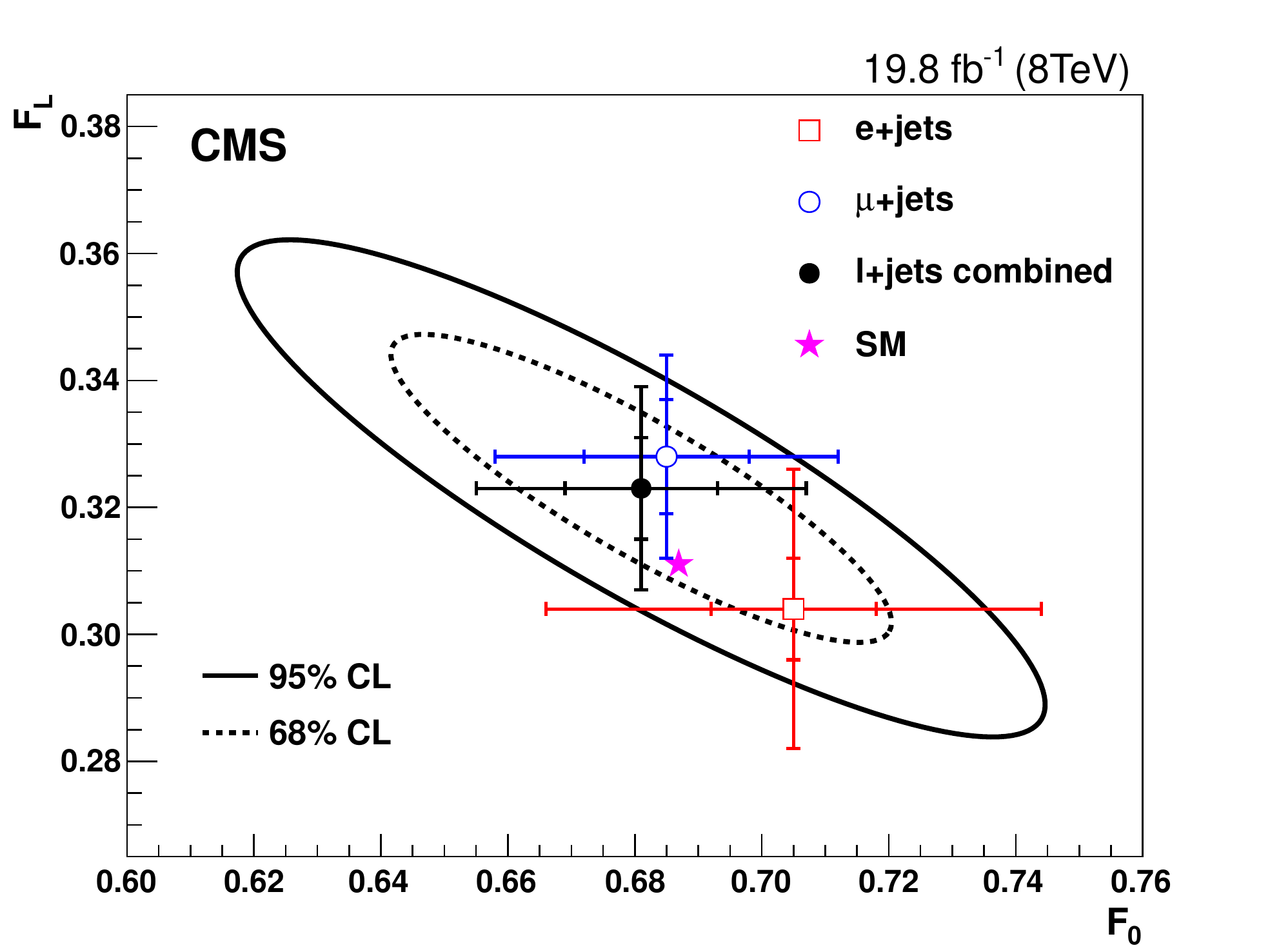}  \includegraphics[scale=0.45]{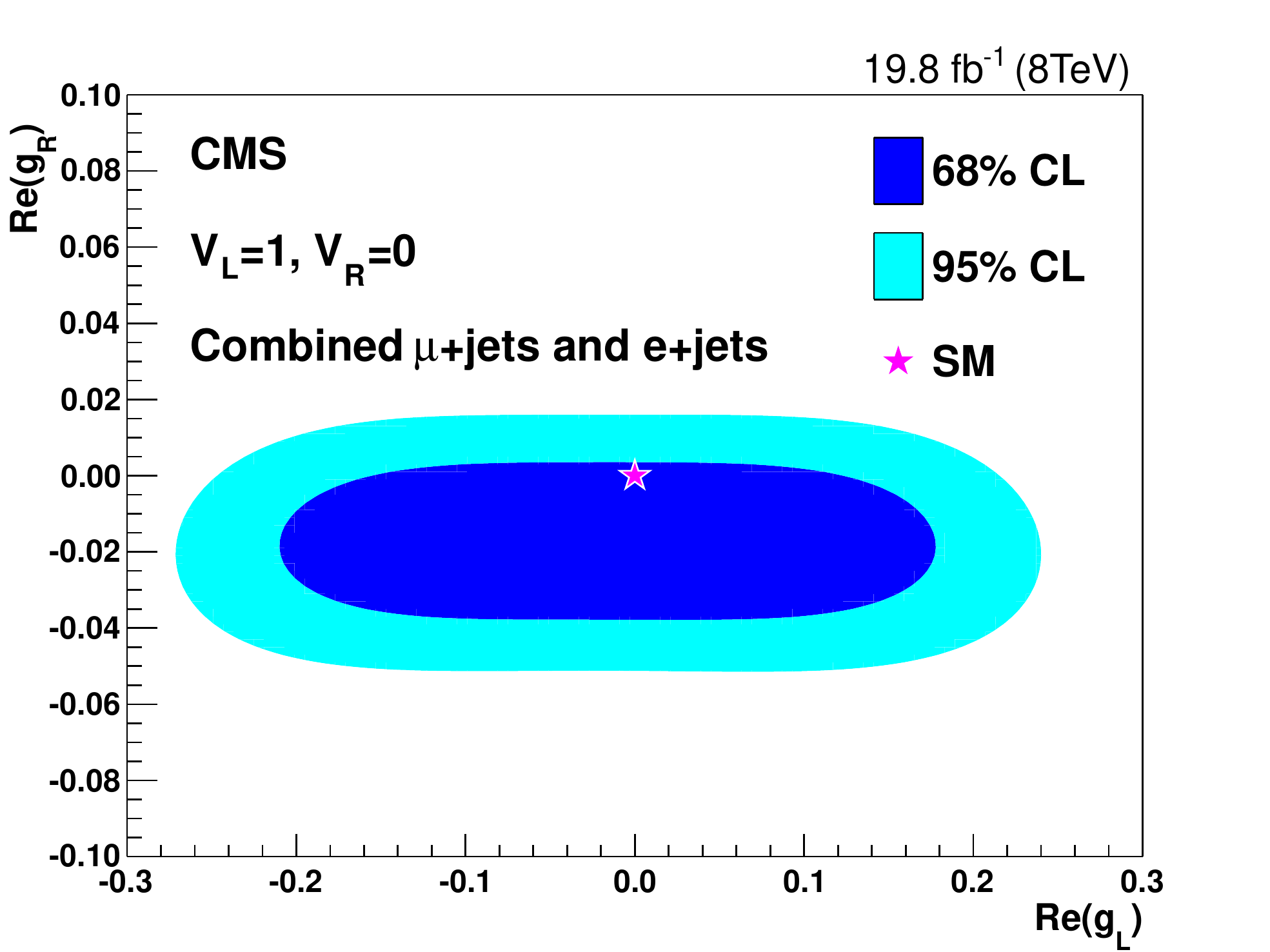} }
\caption{
Left: the measured $\PW$ boson helicity fractions in the $(F_\mathrm{0},F_\mathrm{L})$ plane.
The dashed and solid ellipses enclose the allowed two-dimensional 68\% and 95\% CL regions, for the combined $\ell$+jets measurement, taking into account the correlations on the total (including systematic) uncertainties.
The error bars give the one-dimensional 68\% CL interval for the separate $F_\mathrm{0}$ and $F_\mathrm{L}$ measurements, with the inner-tick (outer-tick) mark representing the statistical (total) uncertainty.
 Right: the corresponding allowed regions for the real components of the anomalous couplings $g_\mathrm{L}$ and $g_\mathrm{R}$ at 68\% and 95\% CL, for $V_\mathrm{L}=1$ and $V_\mathrm{R}=0$.
 A region near $\mathrm{Re}(g_\mathrm{L})=0$ and $\mathrm{Re}(g_\mathrm{R}) \gg 0$, allowed by the fit but excluded by the CMS single top quark production measurement, is omitted.
The SM predictions are shown as stars.
  \label{fig:limanom}}
\end{center}
\end{figure*}

\section{Systematic uncertainties}
Systematic effects which could potentially bias the measurement of the $\PW$ boson helicity fractions have been investigated and their corresponding uncertainties determined, as presented in Table~\ref{tab:system}.

Residual corrections are applied in simulation to the jet energy scale (JES), to account for differences between data and simulation.
The momenta of the jets in simulation are also smeared so that the jet energy resolution  (JER) in simulation agrees with that in data.
These corrections and smearings are propagated into  $\ptvecmiss$ to correct its momentum scale.
The uncertainties~\cite{ref:jets} associated with the  JES and JER corrections
are  also propagated to  $\ptvecmiss$, and the full analysis, including the \ttbar~reconstruction and the resulting measurements of the $\PW$ boson helicity fractions, is repeated.
Scale factors are used to correct the b tagging efficiency in simulation,
where those corrections are shifted by their estimated uncertainties, and the full analysis repeated.
Scale factors are also used to correct leptons for their identification, isolation and trigger efficiencies, which are varied within their uncertainties so as to maximise potential shape variations of the predicted \costh~distributions.

To account for any possible bias of the $\PW$ boson helicity measurements due to uncertainties in the normalisation of simulated backgrounds, the assumed cross section for each sample is varied individually \cite{TOP-11-020}.
An uncertainty of 30\%  is used for the normalisation of DY+jets, single top quark, and $\PW$ boson production in association with light-quark or gluon
jet production.
Since the modelling of the simulated  heavy-flavour content of the
 $\PW$+jets sample is known to be inaccurate, an uncertainty of ~$^{+100\%}_{-50\%}$ is assumed for simulated events involving a $\PW$ boson produced in association with b quark jets.
The impact of the
DY+jets normalisation uncertainty
 in the analysis is small, since  it corresponds to only a few percent of the sample composition.
The normalisation of the  multijet background is estimated from control samples and results in an uncertainty of $^{+50\%}_{-50\%}$ in the $\Pe$+jets channel and $^{+40\%}_{-50\%}$ in the $\mu$+jets channel.
Shape uncertainties on the multijet background templates were investigated by comparing the distributions in several different control regions, both in MC and in data,  and were found to be negligible, compared to the much larger normalisation uncertainties.

Several uncertainties from possible systematic effects related to theoretical modelling of the signal are estimated by replacing the default   \ttbar samples
with alternative \ttbar samples and repeating the entire analysis.
Specifically, for the \MADGRAPH interfaced with PYTHIA event generation,
the default   $m_{\PQt}$ value of 172.5\GeV is shifted up and down by 1\GeV;
the renormalisation and factorisation scales are varied down (up) by a factor of 0.5 (2); the kinematic scale used to match jets to partons (matching threshold) is varied down (up) by factor of 0.5 (2); finally, the parton shower and hadronisation modelling is studied in a \ttbar sample simulated with MC@NLO v3.41 \cite{mcnlo} using the PDF set CTEQ6M and interfaced with \HERWIG v6.520 \cite{herwig}.

Uncertainties in the helicity fractions arising from the limited size of the simulated \ttbar samples are
 taken into account, both in the main analysis and in the determination of the modelling uncertainties.
In the former case, these effects are added as a separate source of uncertainty.
In the latter case, the systematic uncertainties in the $\PW$ boson helicity are assigned to be the larger of either (i) the statistical precision of the limited sample size or (ii) the systematic shift of the central value with respect to the reference \ttbar sample.

  The shape of the \pt spectrum for top quarks, as measured by the differential cross section for top quark pairs \cite{diff1,diff2}, has been found to be softer than the predictions from \MADGRAPH simulation.
The effect of this mismodelling is estimated by reweighting the events in the simulated \ttbar sample, so that
the top quark \pt~at parton level in the MC describes the unfolded data distribution.
Further, the systematic effects due to the PDFs used to simulate the signal and background samples are estimated   according to
 the prescriptions described in \cite{pdf4lhc,pdf4lhc1}, using  NNPDF21~\cite{NNPDF}
and MSTW2008lo68cl~\cite{pdfs} PDF sets as alternatives to those used at  generation.
Finally, uncertainties related to the modelling of the pileup in simulated events are also taken into account.

The total systematic uncertainty is given by the sum in quadrature of all uncertainties described above.

\section{Results}

The measurements of the $\PW$ boson helicity fractions, using \costh~from the leptonic branch of \ttbar~events that decay into $\Pe$+jets or $\mu$+jets, including the full combination of these two measurements, are shown in Table~\ref{tab:results}.
Within an individual channel, the helicity parameters $F_\mathrm{0}$ and $F_\mathrm{L}$ are fit simultaneously, but they are strongly anti-correlated due to the unitarity constraint $F_\mathrm{L}+F_\mathrm{ 0}+F_\mathrm{R}=1$, as indicated by the statistical correlation coefficient $\rho_\mathrm{0,L}$ given in the table.
The separate helicity measurements from the $\Pe$+jets and $\mu$+jets channels are combined into a single $\ell$+jets measurement using the BLUE method \cite{Lyons:1988rp,Valassi:2003mu}, taking into account all uncertainties and their possible correlations.
Uncertainties related to lepton efficiency, multijet background estimations, and statistical uncertainties are considered uncorrelated between the $\Pe$+jets and $\mu$+jets analyses, while all other uncertainties are assumed to be fully correlated.
The combined $\ell$+jets measurement of the helicity fractions is dominated by the $\mu$+jets channel, with
weights more than double those of the $\Pe$+jets channel.
The $\chi^2$ of the combination is 2.13 for 2 degrees of freedom, corresponding to a probability of 34.5\%. The measurement uncertainties are dominated by systematic effects that are correlated between both the $\Pe$+jets and $\mu$+jets channels.
The combined  $F_\mathrm{0}$ and $F_\mathrm{L}$ values are anti-correlated with a statistical correlation coefficient $\rho_\mathrm{0,L}=-0.959$. The total correlation coefficient,
considering both statistical and systematic uncertainties,  is found to be  $-0.870$.
The measured helicity fractions presented in Table~\ref{tab:results} are consistent with the SM predictions given at NNLO accuracy~\cite{Czarnecki:2010gb}.
Figure~\ref{fig2} shows, separately for the $\Pe$+jets and $\mu$+jets channels, the distributions for the cosine of the helicity angles from the leptonic branch,
which are used in the helicity measurements, and the distributions of the corresponding absolute values from the hadronic branch, for comparison purposes.
The simulated samples involving top quarks used in the figure were produced using the measured values for the $\PW$ boson helicity fractions, as determined from the combined $\ell$+jets fit.
Left-handed W bosons tend to populate the region at $\cos\theta^{*}\approx -1$, where the charged lepton overlaps with the b quark.
However, the angular separation requirement between leptons and jets removes most of the events
near $\cos\theta^{*}= -1$.
Very few events are expected in the region preferred by right-handed bosons, near $\cos\theta^{*}= +1$.
However, due to resolution effects,
the reconstructed distribution does not fall as rapidly  as
expected in that region, where the charged lepton and b quark have  opposite directions.
For these reasons, the shape of the reconstructed $\cos\theta^{*}$ distribution differs from that expected in the SM (Fig. 1). These features are well reproduced by the simulation, and taken into account in the measurement.

\begin{table*}[h]
\begin{center}
\topcaption{Measurements of the $\PW$ boson helicity fractions from lepton+jets
final states in \ttbar~decays.
The helicity fractions  $ F_\mathrm{0}$  and $ F_\mathrm{L}$ are measured simultaneously and are strongly anti-correlated, as indicated by a
correlation coefficient $\rho_\mathrm{0,L}$, because $F_\mathrm{R}$ is derived from the unitarity condition.  \label{tab:results}}
\tabcolsep=0.15cm
\ifthenelse{\boolean{cms@external}}{}{\resizebox{\textwidth}{!}}
{
\begin{tabular}{lrrrc} \hline
  Channel &  $F_0$ $\pm$ (stat) $\pm$ (syst) & $F_L$ $\pm$ (stat) $\pm$ (syst)  & $F_R$ $\pm$ (stat) $\pm$ (syst) & $\rho_{0,L}$
\\ \hline
$\Pe$+jets   &   0.705 $\pm$0.013$\pm$0.037    & 0.304$\pm$0.009$\pm$0.020    &  $-0.009\pm 0.005 \pm$0.021  & $-0.950$ \\
 $\mu$+jets  &   0.685 $\pm$0.013$\pm$0.024    & 0.328$\pm$0.009$\pm$0.014    &  $-0.013\pm 0.005 \pm$0.017  & $-0.957$ \\
$\ell$+jets  &   0.681 $\pm$0.012$\pm$0.023    & 0.323$\pm$0.008$\pm$0.014    &  $-0.004\pm 0.005 \pm$0.014  & $-0.959$ \\ \hline
\end{tabular}
}
\end{center}
\end{table*}

Using these results, limits on anomalous couplings are obtained  by fixing the two vector couplings in Eq. (\ref{eq:Wtb0}) to their SM values, $V_\mathrm{L}=1$ and  $V_\mathrm{R}=0$, and choosing the tensor couplings, $\mathrm{Re}(g_\mathrm{L})$ and $\mathrm{Re}(g_\mathrm{R})$, as free parameters.
The combined $\ell$+jets measurement of the $\PW$ boson helicity fractions $F_\mathrm{0}$ and $F_\mathrm{L}$ is reinterpreted in terms of the tensor couplings, $\mathrm{Re}(g_\mathrm{L})$ and $\mathrm{Re}(g_\mathrm{ R})$, using the relationships between the $\PW$ boson helicity fractions  given in Ref.~\cite{SaavedraBernabeu}.

The $\PW$ boson helicity measurements are displayed in the $(F_\mathrm{0},F_\mathrm{L})$ plane in Fig. \ref{fig:limanom} (left), together with their one-dimensional statistical (inner-tick mark) and total (outer-tick mark) error bars.
The full two-dimensional confidence level (CL) contours corresponding to 68\% (dashed line) and 95\% (solid line)
probabilities are also displayed for the combined measurement.
The SM prediction is shown as a  star and lies within the 68\% CL contour.
The corresponding regions in the $(\mathrm{Re}(g_\mathrm{L}),\mathrm{Re}(g_\mathrm{R}))$ plane, allowed at 68\% (dark contour) and 95\% CL (light contour), are shown in
Fig. \ref{fig:limanom} (right),
 together with the SM value.
They are derived from Fig. \ref{fig:limanom} (left), using the relationships between the $\PW$ boson helicity fractions and the anomalous
couplings given in Ref.~\cite{SaavedraBernabeu}.
A region near $\mathrm{Re}(g_\mathrm{L})=0$ and $\mathrm{Re}(g_\mathrm{R}) \gg 0$, allowed by the fit but excluded by the CMS single top quark production measurement \cite{tchann}, is not shown.

If the right-handed component  $F_\mathrm{R}$ is bound to zero,
consistently with the SM within the precision of
the current measurement, the combined $\ell$+jets measurement amounts to
$F_\mathrm{0}=0.661\pm0.006\stat\pm 0.021\syst$.
In this case,  $F_\mathrm{L}$ is obtained via the unitarity constraint and yields  $F_\mathrm{L}=0.339\pm0.006\stat\pm 0.021\syst$.

\section{Summary}

A measurement of  the $\PW$ boson helicity fractions in top quark pair events decaying  in the $\Pe$+jets and $\mu$+jets channels
has been presented, using proton-proton collision data at \mbox{$\sqrt{s}= 8\TeV$,} and corresponding to an integrated luminosity of 19.8\fbinv.
The helicity fractions $F_\mathrm{0}$ and $F_\mathrm{L}$ are measured with a precision of better than 5\%,
yielding the most accurate experimental determination of the $\PW$ boson helicity fractions in \ttbar processes to date.
The measured $\PW$ boson helicity fractions are
 $F_\mathrm{0}=0.681\pm0.012\stat\pm 0.023\syst$,  $F_\mathrm{L}=0.323\pm 0.008\stat\pm 0.014\syst$,
and  $F_\mathrm{R}=-0.004\pm 0.005\stat\pm 0.014\syst$,
with a correlation coefficient of $-0.87$ between  $F_\mathrm{0}$ and  $F_\mathrm{L}$,
and they are consistent with the expectations from the standard model.

\begin{acknowledgments}

\hyphenation{Bundes-ministerium Forschungs-gemeinschaft Forschungs-zentren}

We congratulate our colleagues in the CERN accelerator departments for the excellent performance of the LHC and thank the technical and administrative staffs at CERN and at other CMS institutes for their contributions to the success of the CMS effort. In addition, we gratefully acknowledge the computing centres and personnel of the Worldwide LHC Computing Grid for delivering so effectively the computing infrastructure essential to our analyses. Finally, we acknowledge the enduring support for the construction and operation of the LHC and the CMS detector provided by the following funding agencies: BMWFW and FWF (Austria); FNRS and FWO (Belgium); CNPq, CAPES, FAPERJ, and FAPESP (Brazil); MES (Bulgaria); CERN; CAS, MOST, and NSFC (China); COLCIENCIAS (Colombia); MSES and CSF (Croatia); RPF (Cyprus); MoER, ERC IUT and ERDF (Estonia); Academy of Finland, MEC, and HIP (Finland); CEA and CNRS/IN2P3 (France); BMBF, DFG, and HGF (Germany); GSRT (Greece); OTKA and NIH (Hungary); DAE and DST (India); IPM (Iran); SFI (Ireland); INFN (Italy); MSIP and NRF (Republic of Korea); LAS (Lithuania); MOE and UM (Malaysia); BUAP, CINVESTAV, CONACYT, LNS, SEP, and UASLP-FAI (Mexico); MBIE (New Zealand); PAEC (Pakistan); MSHE and NSC (Poland); FCT (Portugal); JINR (Dubna); MON, RosAtom, RAS and RFBR (Russia); MESTD (Serbia); SEIDI and CPAN (Spain); Swiss Funding Agencies (Switzerland); MST (Taipei); ThEPCenter, IPST, STAR and NSTDA (Thailand); T\"UBITAK and TAEK (Turkey); NASU and SFFR (Ukraine); STFC (United Kingdom); DOE and NSF (USA).

Individuals have received support from the Marie-Curie programme and the European Research Council and EPLANET (European Union); the Leventis Foundation; the Alfred P. Sloan Foundation; the Alexander von Humboldt Foundation; the Belgian Federal Science Policy Office; the Fonds pour la Formation \`a la Recherche dans l'Industrie et dans l'Agriculture (FRIA-Belgium); the Agentschap voor Innovatie door Wetenschap en Technologie (IWT-Belgium); the Ministry of Education, Youth and Sports (MEYS) of the Czech Republic; the Council of Science and Industrial Research, India; the HOMING PLUS programme of the Foundation for Polish Science, cofinanced from European Union, Regional Development Fund; the Mobility Plus programme of the Ministry of Science and Higher Education (Poland); the OPUS programme of the National Science Center (Poland); the Thalis and Aristeia programmes cofinanced by EU-ESF and the Greek NSRF; the National Priorities Research Program by Qatar National Research Fund; the Programa Clar\'in-COFUND del Principado de Asturias; the Rachadapisek Sompot Fund for Postdoctoral Fellowship, Chulalongkorn University (Thailand); the Chulalongkorn Academic into Its 2nd Century Project Advancement Project (Thailand); and the Welch Foundation, contract C-1845.

\end{acknowledgments}

\clearpage

\bibliography{auto_generated}

\cleardoublepage \appendix\section{The CMS Collaboration \label{app:collab}}\begin{sloppypar}\hyphenpenalty=5000\widowpenalty=500\clubpenalty=5000\textbf{Yerevan Physics Institute,  Yerevan,  Armenia}\\*[0pt]
V.~Khachatryan, A.M.~Sirunyan, A.~Tumasyan
\vskip\cmsinstskip
\textbf{Institut f\"{u}r Hochenergiephysik,  Wien,  Austria}\\*[0pt]
W.~Adam, E.~Asilar, T.~Bergauer, J.~Brandstetter, E.~Brondolin, M.~Dragicevic, J.~Er\"{o}, M.~Flechl, M.~Friedl, R.~Fr\"{u}hwirth\cmsAuthorMark{1}, V.M.~Ghete, C.~Hartl, N.~H\"{o}rmann, J.~Hrubec, M.~Jeitler\cmsAuthorMark{1}, A.~K\"{o}nig, I.~Kr\"{a}tschmer, D.~Liko, T.~Matsushita, I.~Mikulec, D.~Rabady, N.~Rad, B.~Rahbaran, H.~Rohringer, J.~Schieck\cmsAuthorMark{1}, J.~Strauss, W.~Treberer-Treberspurg, W.~Waltenberger, C.-E.~Wulz\cmsAuthorMark{1}
% ...
\vskip\cmsinstskip
% ...
1:~~Also at Vienna University of Technology, Vienna, Austria\\
% ...

\end{sloppypar}
\end{document}